\documentclass[a4paper]{jpconf}
\usepackage{graphicx}
\begin{document}
\title{Configuration mixing effects in neutron-rich carbon isotopes}

\author{C.X. Yuan$^{1}$, F.R. Xu$^{2}$*, C. Qi$^{3}$}

\address{$^{1}$ Sino-French Institute of Nuclear Engineering and Technology,
Sun Yat-sen University, Tangjiawan, Zhuhai, Guangdong, 519082,
China}

\address{$^{2}$ State Key Laboratory of Nuclear Physics and Technology,
School of Physics, Peking University, Beijing 100871, China}

\address{$^{3}$ KTH (Royal Institute of Technology), Alba Nova
University Center, SE-10691 Stockholm, Sweden}

\ead{frxu@pku.edu.cn}
\begin{abstract}

Shell model calculations are done to study the structure of
neutron-rich carbon isotopes. For both even-A and odd-A neutron-rich
carbon isotopes, the energy levels are strongly affected by the
configuration mixing of valence neutrons. The calculated energy
levels in the nucleus $^{17}$C are significantly improved compared
with experimental values when the model space of the three valence
neutrons is enlarged from pure $\nu(0d_{5/2})^{3}$ configuration to
full $sd$ space. We also investigate the configuration mixing effect
on the $B(E2)$ values in even-even nuclei $^{16-20}$C.

\end{abstract}

\section{Introduction}

The shell structure evolves when we go from the beta-stability line
to drip lines~\cite{solin2008}. New magic numbers emerge in neutron
or proton-rich nuclei, such as the new $N=14$ and $N=16$ magic
numbers in oxygen isotopes~\cite{janssens2009,stanoiu2004}. The
proton-neutron interaction, especially its tensor component, plays
an important role in the shell
evolution~\cite{solin2008,otsuka2001,otsuka2005,otsuka2010,suzuki2009}.
In case of the $N=14$ shell evolution, from oxygen to carbon
isotopes, two protons are removed from the $0p_{1/2}$ orbit.
$0p_{1/2}$ proton attracts $0d_{5/2}$ neutron more than $1s_{1/2}$
neutron~\cite{yuan2012NPA}. The $N=14$ gap thus becomes smaller in
carbon isotopes than in oxygen isotopes. The attractive
neutron-neutron interaction also contributes to the $N=14$ shell
evolution~\cite{yuan2012NPA}. When the $N=14$ shell gap becomes
smaller, neutrons move from $0d_{5/2}$ to $0s_{1/2}$ orbit. The
attractive neutron-neutron interaction $V^{nn}_{1s_{1/2}1s_{1/2}}$
and $V^{nn}_{0d_{5/2}0d_{5/2}}$ thus make the $N=14$ shell gap even
smaller~\cite{yuan2012NPA}. The $N=14$ shell existing in oxygen
isotopes reduces in nitrogen isotopes~\cite{sohler2008} and
disappears in carbon isotopes~\cite{stanoiu2008} which is caused by
both proton-neutron and neutron-neutron
interaction~\cite{yuan2012NPA}. The configuration mixing between
neutron $0d_{5/2}$ and $1s_{1/2}$ orbits becomes important in carbon
isotopes. In this paper, we will analyze effects caused by
configuration mixing in neutron-rich carbon isotopes.

\section{\label{sec:level2}Shell-model framework}

The model space we used in this paper is the $psd$ model space,
which includes $0p_{3/2}$, $0p_{1/2}$, $0d_{5/2}$, $1s_{1/2}$, and
$0d_{3/2}$ orbits~\cite{brown}. The well-established
WBP~\cite{WBT92}, WBT~\cite{WBT92}, and MK~\cite{MK} effective
Hamiltonians are used. We restrict that maximum zero or two nucleons
can be excited from $p$ to $sd$ shell, which are denoted as
$0\hbar\omega$ and $2\hbar\omega$, respectively. The calculations
are carried out with a newly-established parallel shell-model code
described in Ref.~\cite{Qi07}

In the case of $2\hbar\omega$ calculations, the spurious states
caused by the center-of-mass motion need to be removed. The shell
model Hamiltonian $H$ can be modified to be $H'=H+\beta H_{c.m.}$
where $H_{c.m.}$ is the center-of-mass Hamiltonian~\cite{cm1974}.
The center-of-mass excitations can be moved to high excitation
energies by setting a large positive $\beta$ value. Thus the
spurious states do not appear at low excitation energy. We use
$\beta=100$~MeV in the present work, the same as in previous
works~\cite{umeya20082,ma2010,utsuno2011}.

\section{\label{sec:level3}Calculations and discussions}

\begin{figure}
\begin{center}
\includegraphics[width=20pc]{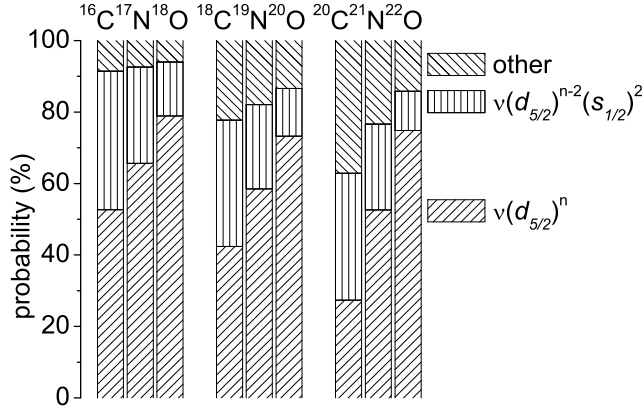}
\caption{\label{even}Probabilities of each configuration in the
ground states of neutron-rich carbon, nitrogen, and oxygen isotopes
calculated with WBP interaction. $n$ indicates the number of valence
neutrons in $sd$ shell. $n=2,~4,~6$ correspond to $^{16}$C,
$^{18}$C, and $^{20}$C, respectively. }
\end{center}
\end{figure}

\begin{figure}
\begin{center}
\includegraphics[width=12pc]{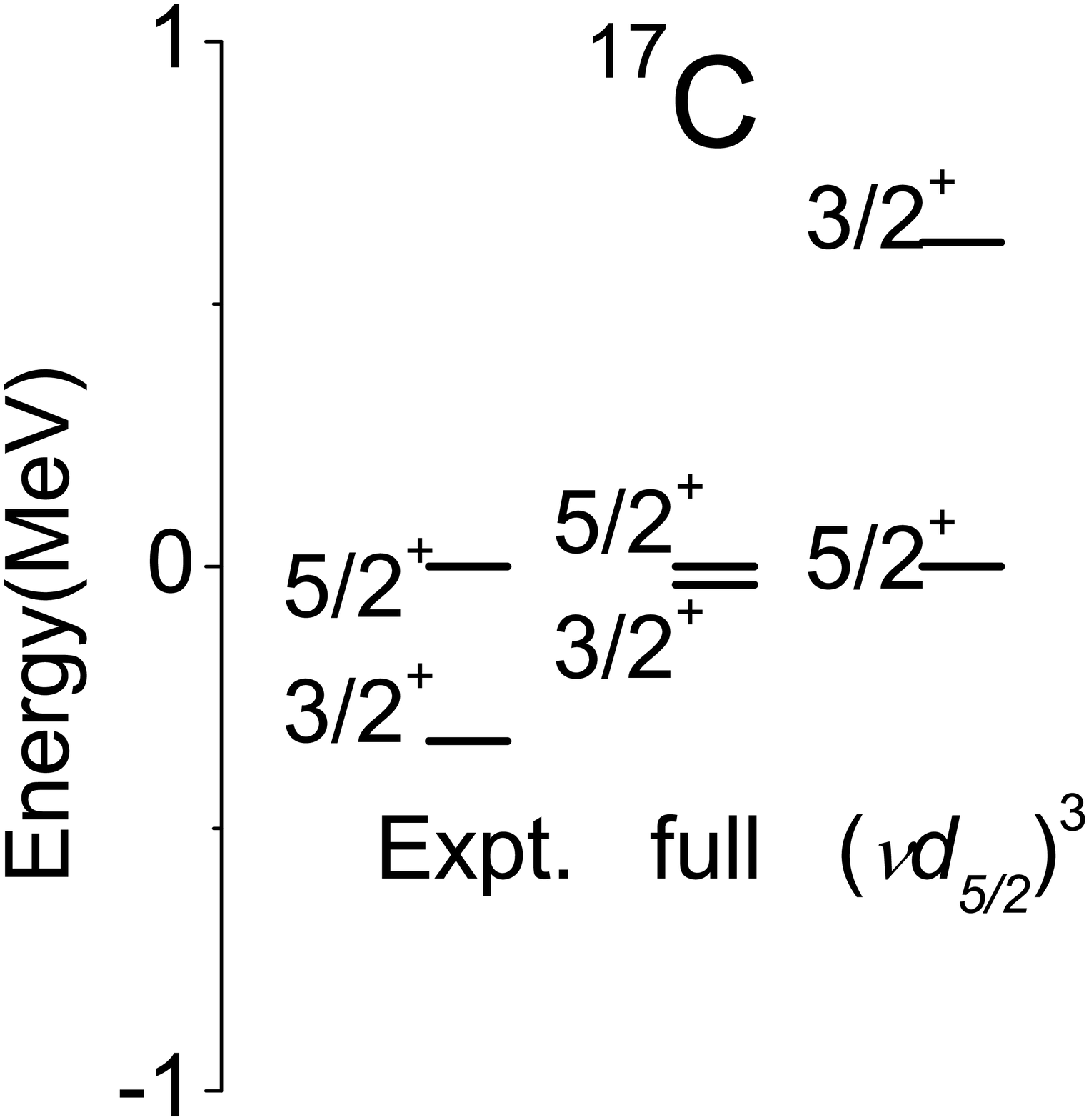}
\includegraphics[width=12pc]{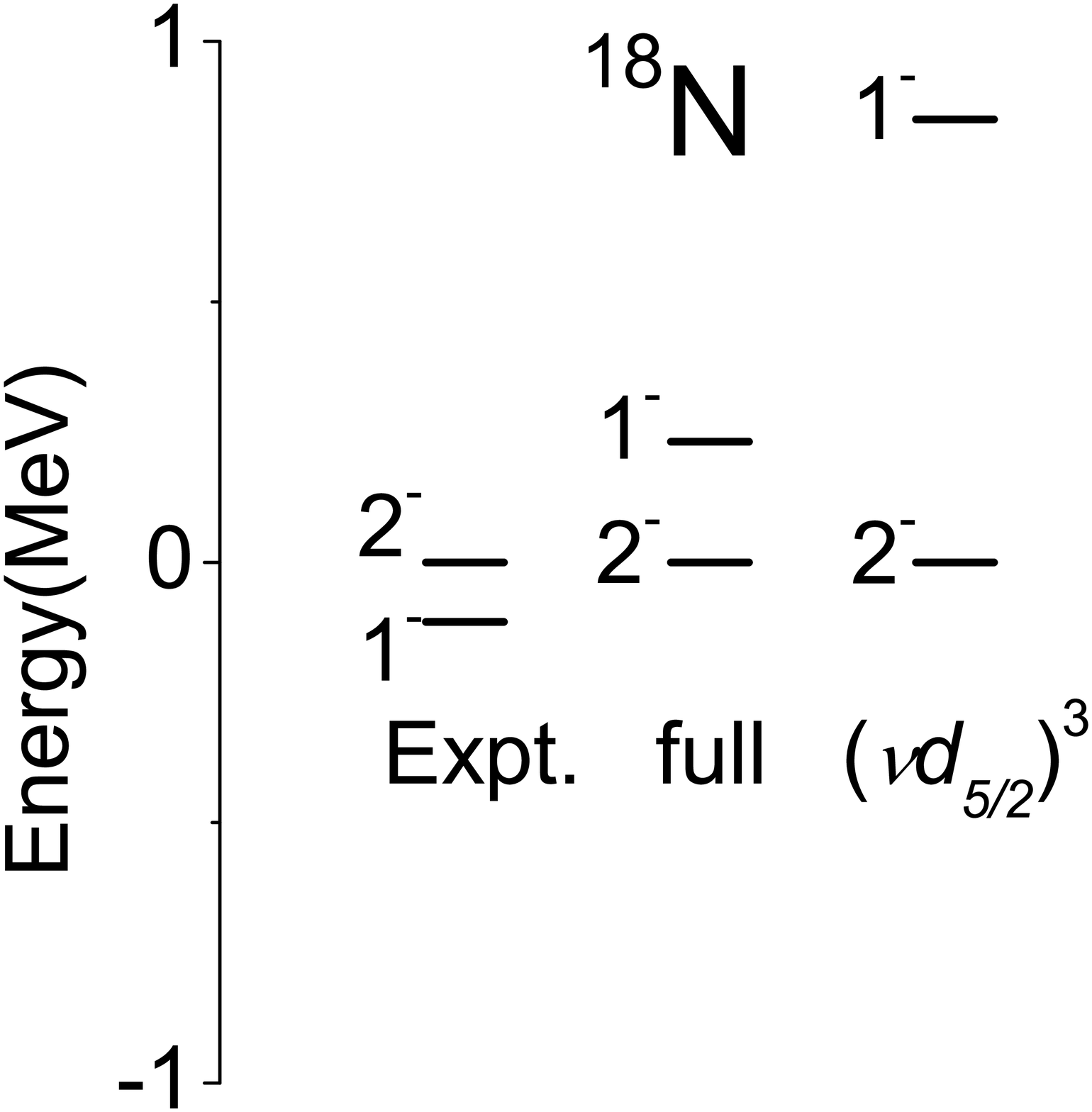}
\includegraphics[width=12pc]{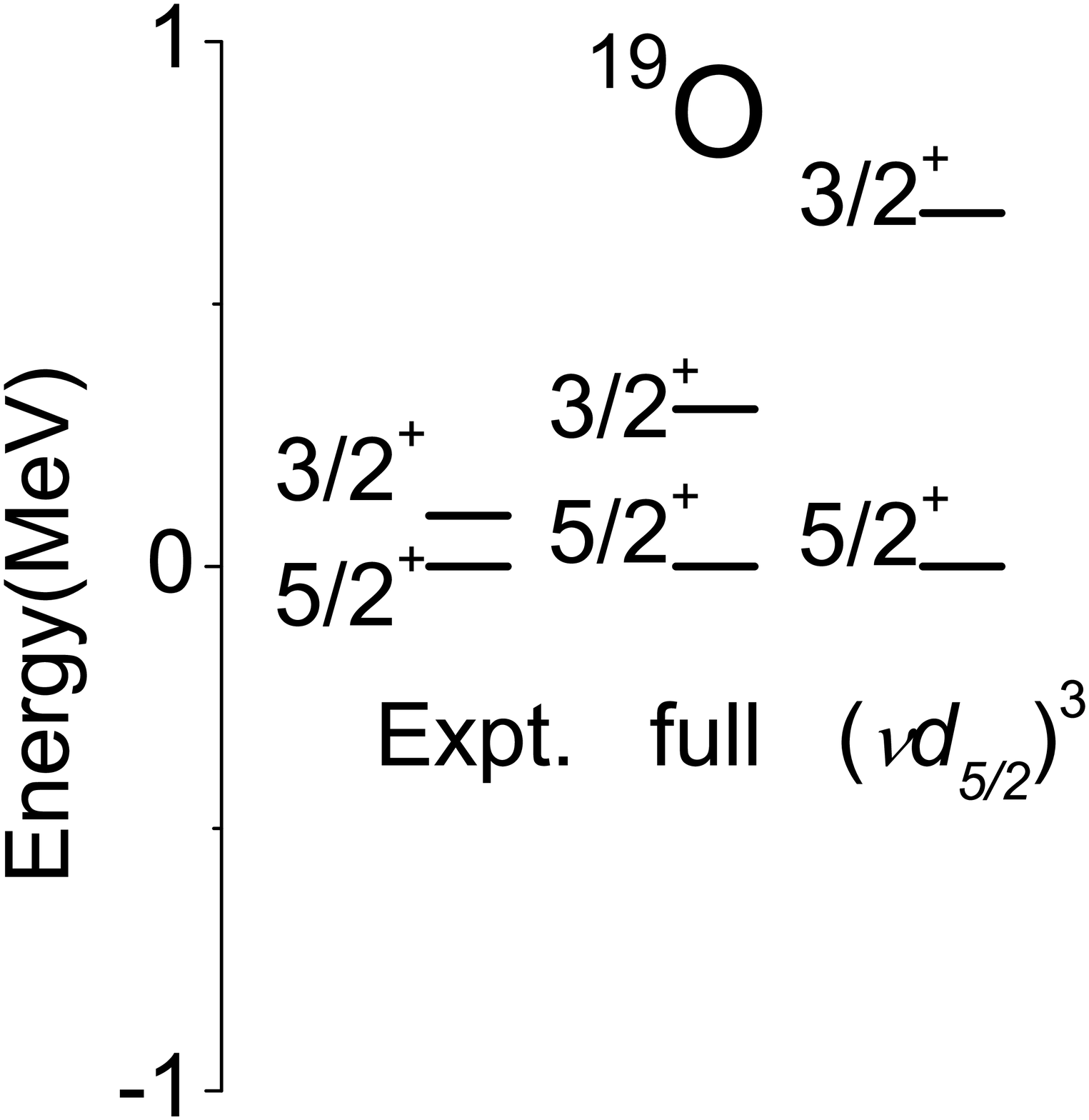}
\caption{\label{odd1}Energy levels of $N=11$ isotones. Experimental
values~\cite{stanoiu2004,sohler2008,stanoiu2008} are compared with
WBP results with two model space for valence neutron. One is full
$sd$ model space, the other is $\nu(0d_{5/2})^{3}$ which means the
three valence neutrons are restricted on $0d_{5/2}$ orbit.}
\end{center}
\end{figure}

The $E(2^{+}_{1})$ in $N=14$ isotones decreases quickly from oxygen
to carbon isotopes which indicates that the $N=14$ shell gap
existing in oxygen isotopes disappears in carbon
isotopes~\cite{stanoiu2004,stanoiu2008}. The excitation energies in
carbon isotopes are strongly affected by the configuration mixing
between neutron $0d_{5/2}$ and $1s_{1/2}$ orbit. In $^{18}$O,
$^{20}$O, and $^{22}$O, the valence neutrons which occupy $0d_{5/2}$
orbit couple to $J=0$ in ground states. When excited, the
$2^{+}_{1}$ states gain energy by the valence neutrons coupling to
$J=2$ in $^{18}$O and $^{20}$O. In $^{22}$O of which valence
neutrons have fully occupied the $0d_{5/2}$ orbit, the
$E(2^{+}_{1})$ is gained by exciting a neutron from $0d_{5/2}$ to
$1s_{1/2}$ orbit. In case of C isotopes, the $N=14$ shell is
disappeared. So the valence neutrons in C isotopes are not only
occupy $\nu0d_{5/2}$ orbit but also $\nu1s_{1/2}$ orbit
~\cite{yuan2012NPA}. The $2^{+}_{1}$ states of $^{16}$C, $^{18}$C
and $^{20}$C are excited by both neutrons coupling to $J=2$ in
$0d_{5/2}$ and a neutron moving to $1s_{1/2}$ orbit. This simple
consideration is supported by shell-model calculations, as shown in
Fig.~\ref{even}. The neutron-rich oxygen isotopes contain very large
(around $80\%$) pure $\nu0d_{5/2}$ configuration. On the other hand,
the strong configuration mixing, especially the mixing between
$\upsilon0d_{5/2}$ and $\nu1s_{1/2}$ orbits, exists in neutron-rich
carbon isotopes.

The energy levels of odd-A neutron-rich carbon isotopes also show
the effects of configuration mixing between $\nu0d_{5/2}$ and
$\nu1s_{1/2}$ orbits. The energy levels of $N=11$ isotones are hard
to be exactly described by shell model because of the strong
configuration mixing~\cite{wiedeking20082}. A recent suggested
Hamiltonian can well reproduce the energy levels of $^{17}$C,
$^{18}$N, and $^{19}$O~\cite{yuan2012PRC}. Figure~\ref{odd1} shows
how the configuration mixing drive the energy levels of $^{17}$C,
$^{18}$N, and $^{19}$O. Here we present the $0\hbar\omega$ results,
as the $2\hbar\omega$ results do not show much difference. In the
case of $0\hbar\omega$ calculations, the three valence neutrons in
$^{17}$C, $^{18}$N, and $^{19}$O can be active in three $sd$ orbits.
When we enlarge the model space of valence neutrons from pure
$\nu(0d_{5/2})^{3}$ to full $sd$ configuration, the energy levels of
these nuclei can be much improved compared with experimental values,
especially for $^{17}$C.

\begin{figure}
\begin{center}
\includegraphics[width=24pc]{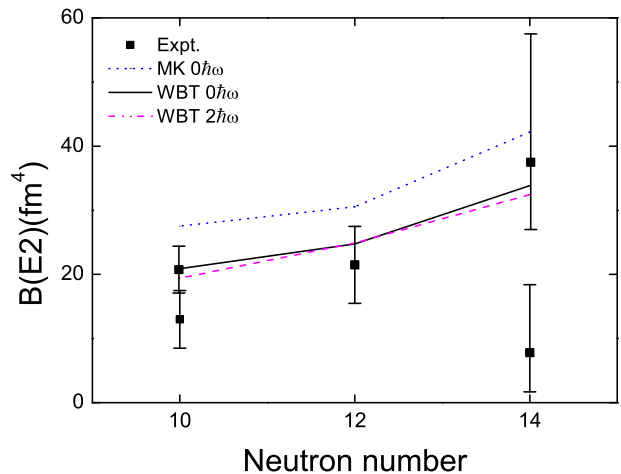}
\caption{\label{BE2}Calculated B(E2) values for even-even
$^{16-20}$C. Experimental values are taken from
Refs.~\cite{petri2011,elekes2009,wiedeking2008,ong2008}. }
\end{center}
\end{figure}

\begin{figure}
\begin{center}
\includegraphics[width=24pc]{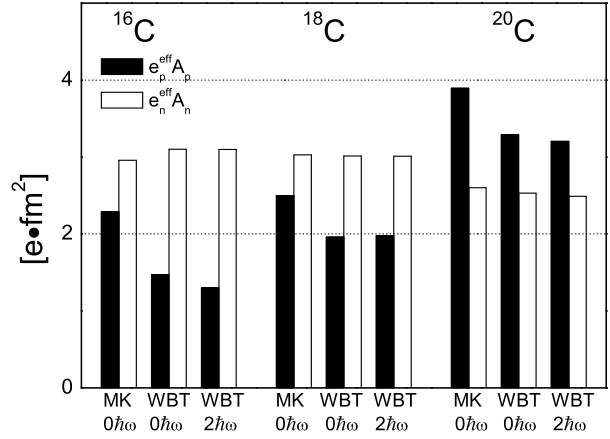}
\caption{\label{eA}Calculated E2 transition matrix elements for
even-even $^{16-20}$C.}
\end{center}
\end{figure}

\begin{figure}
\begin{center}
\includegraphics[width=20pc]{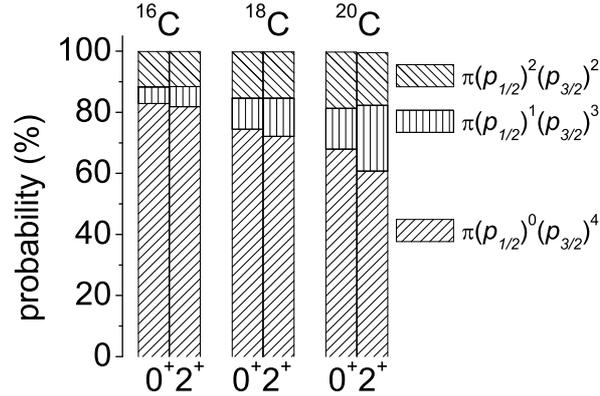}
\caption{\label{proton}Calculated WBT result of proton
configurations for even-even $^{16-20}$C.}
\end{center}
\end{figure}

Compared with valence neutrons in neutron-rich carbon isotopes, the
valence protons in such nuclei are less active as the $Z=6$ gap is
much larger than the $N=14$ gap. However, the configuration mixing
between valence protons strongly affects the $E2$ transition in
neutron-rich carbon isotopes~\cite{yuan2012NPA}.

The shell-model $B(E2)$ results agree well with the observed value
in neutron rich C isotopes~\cite{petri2011}, also as show in
Fig.~\ref{BE2}. In our calculations, an approximate $1/A$ dependence
effective charge is used~\cite{sagawa2004}. The very small $B(E2)$
value is from an inelastic scattering measurement~\cite{elekes2009}.
To clarify the contribution from valence neutrons and protons, we
present the $e_{p(n)}^{eff}A_{p(n)}$ in Fig.~\ref{eA}. Both proton
and neutron contribute a lot in $B(E2)$ values. There is a sudden
increase of $e_{p}^{eff}A_{p}$ at $^{20}$C when the
$e_{n}^{eff}A_{n}$ changes little from $^{16}$C to $^{20}$C. As
discussed in Ref.~\cite{petri2011}, the reduce proton
$\pi0p_{1/2}-\pi0p_{3/2}$ gap from $^{16}$C to $^{20}$C is a reason
that the proton excitation enlarges from $^{16}$C to $^{20}$C. When
the neutron number increasing, the $\nu0d_{5/2}$ attracts
$\pi0p_{1/2}$ orbit more than $\pi0p_{3/2}$ orbit both in WBT and MK
interaction. The energy gap between $\pi0p_{1/2}$ and $\pi0p_{3/2}$
decreases from $^{16}$C to $^{20}$C.

We present in Fig.~\ref{proton} the proton configuration of
$^{16}$C, $^{18}$C and $^{20}$C calculated with WBT interaction.
From $^{16}$C to $^{20}$C, the $(\pi0p_{1/2})^{0}$ configuration
reduces and $(\pi0p_{1/2})^{1}$ configuration increases. The $Ap$ is
most contributed by the transition from $(\pi0p_{1/2})^{0}$ in
ground state to $(\pi0p_{1/2})^{1}$ in $2_{1}^{+}$ state and from
$(\pi0p_{1/2})^{1}$ in ground state to $(\pi0p_{1/2})^{0}$ in
$2_{1}^{+}$ state. These two proton transitions increase from
$^{16}$C to $^{20}$C. This simple analysis does not consider the
neutron configurations. Both $(\pi0p_{1/2})^{0}$ and
$(\pi0p_{1/2})^{1}$ relate to many neutron configurations.
Considering all possible configurations, $e_{p}^{eff}A_{p}$ of
$^{20}$C is $1.5$ times more than that of $^{18}$C as shown in
Fig.~\ref{eA}. In calculation of $B(E2)$ values, this
$e_{p}^{eff}A_{p}$ is squared and lead to the large $B(E2)$ value in
$^{20}$C.

\section{\label{sec:level4}Summary}

The effects of configuration mixing on neutron-rich carbon isotopes
are analyzed within the frame work of shell model. The energy levels
of neutron-rich carbon isotopes are strongly affected by the
configuration mixing of valence neutrons. The configuration mixing
of valence protons influence little on the energy levels. However it
contributes significantly on the large $B(E2)$ values in $^{20}$C.

\section{\label{sec:level5}Acknowledgement}
This work has been supported by the National Natural Science
Foundation of China under Grant Nos. 10975006 and 11235001, and the
Swedish Research Council (VR) under grant No. 621-2010-4723.

\end{document}